# Continuous Molecular Fields Approach Applied to Structure-Activity Modeling

**Igor I. Baskin[1] and Nelly I. Zhokhova[2]**

**Abstract**. The Method of Continuous Molecular Fields is a universal approach to predict various properties of chemical compounds, in which molecules are represented by means of continuous fields (such as electrostatic, steric, electron density functions, etc). The essence of the proposed approach consists in performing statistical analysis of functional molecular data by means of joint application of kernel machine learning methods and special kernels which compare molecules by computing overlap integrals of their molecular fields. This approach is an alternative to traditional methods of building 3D "structure-activity" and "structure-property" models based on the use of fixed sets of molecular descriptors. The methodology of the approach is described in this chapter, followed by its application to building regression 3D-QSAR models and conducting virtual screening based on one-class classification models. The main directions of the further development of this approach are outlined at the end of the chapter.

## 0.1 Introduction

Currently, the leading role in predicting biological activity and physicochemical properties of chemical compounds belongs to methods of chemoinformatics [1-3], which are based on revealing "structure-activity" and "structure-property" relationships using modern statistical, data mining, machine learning and artificial intelligence approaches. Rapid progress of these techniques requires the develop-ment of new tools of machine learning and data mining specially adapted to work with molecular structures of chemical compounds [4].

---

[1] Faculty of Physics, M. V. Lomonosov Moscow State University, *Moscow, 119991, Russia;* e-mail: igbaskin@gmail.com; phone: +7(095)9392677

[2] Faculty of Physics, M. V. Lomonosov Moscow State University, *Moscow, 119991, Russia;* e-mail: zhokhovann@gmail.com; phone: +7(095)9392677

In this section, we consider a new approach to building "structure-activity" and “structure-property” models based on the use of continuous functions on space coordinates (called hereinafter *continuous molecular fields*) to represent molecular structures. Different types of molecular fields can be used for this purpose, including electrostatic, steric, hydrophobic, hydrogen bond donor and acceptor fields, etc. This way to describe chemical structures corresponds well to the physical nature of the molecules, which interact with the environment through molecular fields. The suggested approach is an alternative to traditional methods of representing chemical structures in SAR/QSAR/QSPR (Structure-Activity Relationships / Quantitative Structure-Activity Relationships / Quantitative Structure-Property Relationships) studies by means of fixed-sized vectors of descriptors derived from topological molecular graphs, as well as interaction energies with certain probes calculated at specific points in space (e.g., at the nodes of a hypothetic lattice).

So far, the direct use of continuous molecular fields in their functional form in statistical analysis was not possible because standard data analysis procedures can only work with finite and fixed number of features (molecular descriptors). Only recently, thanks to the development of the statistical learning theory [5] and the methodology of using kernels [6] in machine learning instead of fixed-sized feature vectors, it has become possible to process data of any form and complexity.

*The essence of the Continuous Molecular Fields (CMF) approach consists in performing statistical analysis of functional molecular data by means of joint application of kernel machine learning methods and special kernels which compare molecules by computing overlap integrals of their molecular fields* [7].

The principal novelty of our approach is the ability to conduct a statistical analysis of chemical data represented in the form of continuous molecular fields, i.e. an infinite number of attributes organized in a functional form. In this case, statistical model is not a function relating the values of some properties of chemical compounds with the values of several molecular descriptors, like in the case of traditional SAR/QSAR/QSPR models, but a functional relating the properties of chemical compounds with functions describing spatial distribution of molecular fields. As a result, the resulting models are characterized by continuous fields of model coefficients, which can be visualized in the same manner as molecular fields themselves to provide intuitive interpretation of the models and a deep insight into the nature of the corresponding chemical phenomenon. The principal advantage of this approach follows from a natural, accurate and comprehensive representation of molecules by means of continuous molecular fields.

In combination with kernel based machine learning methods, such as the Support Vector Machines (SVM) [8], the Kernel Ridge Regression (KRR) [9], the Gaussian Processes (GP) [10], etc., continuous molecular fields can be used to provide the quantitative prediction of various properties of chemical compounds. The feasibility of this was demonstrated by us earlier [7]. In conjunction with ker-nel-based methods of one-class classification (novelty detection), such as the One-Class Support Vector Machines (1-SVM) [11], the use of continuous molecular fields allows one to set up a ligand-based virtual screening procedure for searching huge databases of available chemical compounds and retrieving from them potentially biological active compounds (hits), molecular fields of which are similar to the "idealized configuration" (which reflects the structure of the corresponding binding site in biological macromolecule and can be viewed as a "negative image" of its molecular fields) learned by the corresponding one-class model. The feasibility of this approach in practice has also been demonstrated by us [12,13].

## 0.2 Method of Continuous Molecular Fields

3D-QSAR approaches, in which information concerning the spatial structure of molecules is explicitly taken into account, play very important role in medicinal chemistry and drug design [14-16]. Most of them are based on the use of the molecular fields reflecting different types of intermolecular interactions in which molecules under study can be involved. Historically the first and still the most popular 3D-QSAR method is CoMFA (Comparative Molecular Field Analysis) [17], in which electrostatic interactions are approximated by means of the Coulomb law with point partial charges computed for each atom, whereas steric interactions are expressed using the Lennard-Jones potentials with standard force field parameters. Several other types of molecular fields, such as the hydrophobic [18], hydrogen bond donor and acceptor fields [19], molecular orbital fields [20], E-state fields [21], fields of atom-based indicator variables [22] are also used in the framework of the CoMFA method. In the CoMSIA (Comparative Molecular Similarity Indices Analysis) approach [23] the same types of molecular fields are approximated using the Gaussian radial basis functions. In the GRID method [24] the values of various types of molecular fields are computed as interaction energies with certain probe atom or group of atoms placed at grid nodes.

In all of the above approaches molecular fields are evaluated at node points of some imaginary grid surrounding the set of aligned molecules. The advantage of using such lattices lies in the possibility to use the values of molecular field potentials calculated at grid nodes as a vector of descriptors, which can further be fed to some standard statistical analysis procedure, usually PLS (Partial Least Squares) [25], in order to build regression QSAR models. Another appealing feature of this approach is the possibility to visualize the resulting regression coefficients using easily interpretable isosurfaces (usually colored according to the sign of coefficients and the type of the corresponding molecular field) surrounding the molecules. Nonetheless, this approach has certain considerable drawbacks. Indeed, it is necessary to: (1) choose biologically active conformation for each molecule; (2) align in space the training set of molecules; (3) build a lattice around such set of molecules; (4) choose molecular fields and compute their potentials at grid points;

(5) build regression QSAR models. The problems associated with each of these stages are well known and present a challenge for the current stage of the development of the 3D-QSAR methodology.

The CMF approach addresses the problems caused by the necessity to choose a grid of points around molecules. It is known that 3D-QSAR models sharply depend on the spatial orientation and extent of such grid, as well as on the step size (i.e. the distance between the closest points) in it [15]. Another problem caused by the use of grids is very high dimensionality of the regression task caused by the big number of grid points, which precludes the use of many efficient statistical methods. Unfortunately, decrease of the number of grid points through increase of the step size (i.e. the use of coarser grid) or decrease of its extent causes the loss of important information. In order to tackle the problem, we suggest not using grids or fixed probe positions in 3D-QSAR studies. Instead of computing descriptor values at a discrete set of points, we propose to work directly with descriptions of molecular fields in the form of continuous functions on radius vector **r**, by means of specially constructed kernels. The use of such continuous molecular fields seems to be much more natural and corresponding to the real physical picture of the world than the application of certain speculative grids arbitrary chosen to approximate such fields.

It should be mentioned that continuous molecular fields have already been used in QSAR studies. So, indexes of R. Carbó-Dorca, which were used in certain QSAR studies [26], can also be considered as a particular case of continuous molecular fields.

The method of Continuous Molecular Fields (CMF) performs statistical analysis of functional molecular data by means of joint application of kernel machine learning methods and special kernels which compare molecules by computing overlap integrals of their molecular fields [7].

### *0.2.1 Procedure of Kernel Calculation*

The principal element of the CMF approach is the procedure of calculating molecular field kernels. The joint molecular field kernel $K(M_i, M_j)$ that describes the similarity between all molecular fields of molecules $M_i$ and $M_j$ can be calculated as a linear combination of kernels corresponding to each of $N_f$ types of molecular fields:

$$K(M_i, M_j) = \sum_{f=1}^{N_f} h_f K_f(M_i, M_j), \quad (0.1)$$

where $h_f$ is the mixing coefficient of molecular fields; $K_f(M_i, M_j)$ is the kernel describing the similarity between the molecular fields of the *f*th type for the *i*th and $j$th molecules. Function $K(M_i, M_j)$ represents correctly constructed kernel, because a linear combination of kernels is a kernel.

In some cases, we also use the normalized version of the kernels:

$$K_f^{'}(M_i, M_j) = \frac{K_f(M_i, M_j)}{\sqrt{K_f(M_i, M_i) \cdot K_f(M_j, M_j)}} \tag{0.2}$$

The molecular field kernel for each the $f$th type of molecular field is calculated in the framework of the CMF approach by summation of the kernels for each pair of atoms for the $i$th and $j$th molecules:

$$K_f(M_i, M_j) = \sum_{l=1}^{N_i} \sum_{m=1}^{N_j} k_f(A_{il}, A_{jm}), \tag{0.3}$$

where $k_f(A_{il}, A_{jm})$ is the kernel that describes the similarity between the field of the $f$th type of the $l$th atom in the $i$th molecule and $m$th atom in the $j$th molecule; $N_i$ is the number of atoms in the $i$th molecule; $N_j$ is the number of atoms in the $j$th molecule. $K_f(M_i, M_j)$ can be considered as a 1-tuple convolution kernel corresponding to decomposition of a molecule into atoms. The value of kernel $k_f(A_{il}, A_{jm})$ can be calculated by integration of the product of the fields for a pair of atoms over the entire physical space:

$$k_f(A_{il}, A_{jm}) = \iiint \rho_{fil}(\mathbf{r}) \rho_{fjm}(\mathbf{r}) d^3\mathbf{r} \tag{0.4}$$

where $\rho_{fil}(\mathbf{r})$ is the value of molecular field of the $f$th type induced by the $l$th atom of the $i$th molecule at the point $\mathbf{r}$ of the physical space; the $\rho_{fjm}(\mathbf{r})$ is the same magnitude for the $m$th atom of the $j$th molecule. To simplify the integration, one can approximate any molecular field as a weighted sum of Gaussian basis functions. We have found empirically that in most of cases it is sufficient to use a single Gaussian function to represent any kind of fields produced by a single atom, exactly like in the CoMSIA method [23]:

$$\rho_{fil}(\mathbf{r}) = w_{fil} \exp(-\frac{1}{2} \alpha_f \|\mathbf{r} - \mathbf{r}_{il}\|^2), \tag{0.5}$$

where: $\mathbf{r}_{il}$ is the location of the $l$th atom of the $i$th molecule in the physical space; $\alpha_f$ is the fitting parameter for molecular field of the $f$th type; $w_{fil}$ is the weight of the contribution of $l$th atom of the $i$th molecule to the molecular field of the $f$th type. For example, in the case of electrostatic field the $w_{fil}$ is the partial charge on the $l$th atom of the $i$th molecule, for the steric field – the Lennard-Jones potential parameters, for the hydrophobic field – the contribution of a given atom to the to-tal hydrophobicity. Evidently different sets of the values $w_f$ constitute different parameterizations of the CMF approach.

Due to the afore-mentioned approximation, the foregoing integral can be calculated analytically:

$$k_f(A_{il}, A_{jm}) = \iiint \rho_{fil}(\mathbf{r})\rho_{fjm}(\mathbf{r})d^3\mathbf{r} = w_{fil} w_{fjm} \sqrt{\frac{\pi^3}{(\alpha_f)^3}} \exp(-\frac{\alpha_f}{4}\|\mathbf{r}_{il} - \mathbf{r}_{jm}\|^2) \quad (0.6)$$

It should be pointed out that the CMF approach is not confined to the simplest approximations introduced by Eq. 0.5. Any number of Gaussian functions as well as any other sets of basic functions (such as splines, wavelets, etc) can be used for approximating continuous molecular fields. This provides the ability to use complex types of molecular fields, including those derived from quantum chemistry, such as electron density functions.

### *0.2.2 The Use of Continuous Molecular Fields in Conjunction with Regression Kernel-Based Machine Learning Methods*

Kernel $K(M_i,M_j)$, or its normalized version $K'(M_i,M_j)$, can be plugged in any kernel-based machine learning method (such as Support Vector Machine, Kernel Ridge Regression [9], Kernel Partial Least Squares [27], Gaussian Processes, etc.) in order to build regression, classification, novelty detection models, etc.

In the case of 3D-QSAR kernel-based regression models, the value of the predicted property $y_t$ for a new molecule $M_t$ can be calculated using the following expression:

$$y_t = \sum_{j=1}^{N_m} a_j K(M_t, M_j) + b, \quad (0.7)$$

where $N_m$ is the number of molecules in the training set. If Support Vector Regression (SVR) is used for deriving the values of $a_j$ and $b$, the vector $a_j$ appears to be sparse, with non-zero values corresponding to a certain subset of compounds from the training set. In contrast, contributions of all molecules from the training set are needed to make predictions based on regression models built using the Kernel Ridge Regression (KRR) [9], the Kernel Partial Least Squares [27] or the Gaussian Processes machine learning methods.

In addition to the set of adjustable coefficients $a_j$ and $b$ contained in Eq. 0.7, the method CMF also requires calculation of a certain number of adjustable parameters. Among them are the parameter $\nu$ for the support vector regression method $\nu$-SVR and the ridge parameter $\gamma$ for KRR. Their values should be optimized with the aim to improve the predictive capability of the model constructed. In addition, for each molecular field one can adjust the values of up to two parameters: $\alpha_k$ (the rate of molecular field attenuation, that is related to the width of the Gaussian function) and $h_f$ (mixing coefficient, which has the meaning of the relative contribution of molecular field of the *f*th type).

### *0.2.3 Fields of Model Coefficients*

Eq. 0.7 represents the dual form of the regression model, since in it the activity $y_t$ is predicted by considering similarity measures of a test compound $M_t$ in relation to the training set compounds $M_j$. In order to obtain the traditional primal form of the 3D-QSAR model, which involves an explicit consideration of molecular descriptors and regression coefficients, one can make the substitution of Eqs. 0.1 and 0.3-0.6 to Eq. 0.7 to obtain:

$$y_t = \sum_{f=1}^{N_f} h_f \iiint C_f(\mathbf{r}) X_f(\mathbf{r}) d^3\mathbf{r} + b, \qquad (0.8)$$

$$X_f(\mathbf{r}) = \sum_{l=1}^{N_i} \rho_{ftl}(\mathbf{r}) = \sum_{l=1}^{N_i} w_{ftl} \exp(-\frac{\alpha_f}{2} \|\mathbf{r} - \mathbf{r}_{tl}\|^2), \qquad (0.9)$$

$$C_f(\mathbf{r}) = \sum_{j=1}^{N_m} a_j \sum_{m=1}^{N_j} \rho_{fjm}(\mathbf{r}) = \sum_{j=1}^{N_m} a_j \sum_{m=1}^{N_j} w_{fjm} \exp(-\frac{\alpha_f}{2} \|\mathbf{r} - \mathbf{r}_{jm}\|^2) \qquad (0.10)$$

The primal form of the regression model is expressed by Eq. 0.8, in which molecular field $X(\mathbf{r})$ represents the continuous field of molecular descriptors for the test molecule *t*, whereas $C_f(\mathbf{r})$ is a continuous field of the corresponding regression coefficients. The principal distinction of the CMF Eq. (0.8) from that of an ordinary 3D-QSAR linear model lies in the infinite number of point descriptors in the CMF model. As a result, continuous field of molecular field descriptors is used in it instead of several thousands of descriptors computed in CoMFA or CoMSIA at lattice points, continuous field of regression coefficients is used instead of several thousands of regression coefficients obtained by means of the PLS regression, and integration over the entire physical space substitutes summation over the grid points. It also follows from this analysis that iso-surfaces of the fields of regression coefficients $C_f(\mathbf{r})$ could be used in the same manner and for the same purposes as CoMFA and CoMSIA contour maps. Table 0.1 lists four types of molecular fields, shows isosurfaces of KRR regression coefficients for the case of thrombin inhibitors (2-amidinophenylalanines), and also provides chemical interpretation.

**Table 0.1.** Types of molecular fields, isosurfaces of KRR regression coefficients and chemical interpretation for the case of thrombin inhibitors (2-amidinophenylalanines). In all cases the isosurfaces are superimposed over the structure of one of inhibitors.

| Type of molecular field | Isosurfaces of the fields of regression coefficients | Chemical interpretation |
|---|---|---|
| Electrostatic | 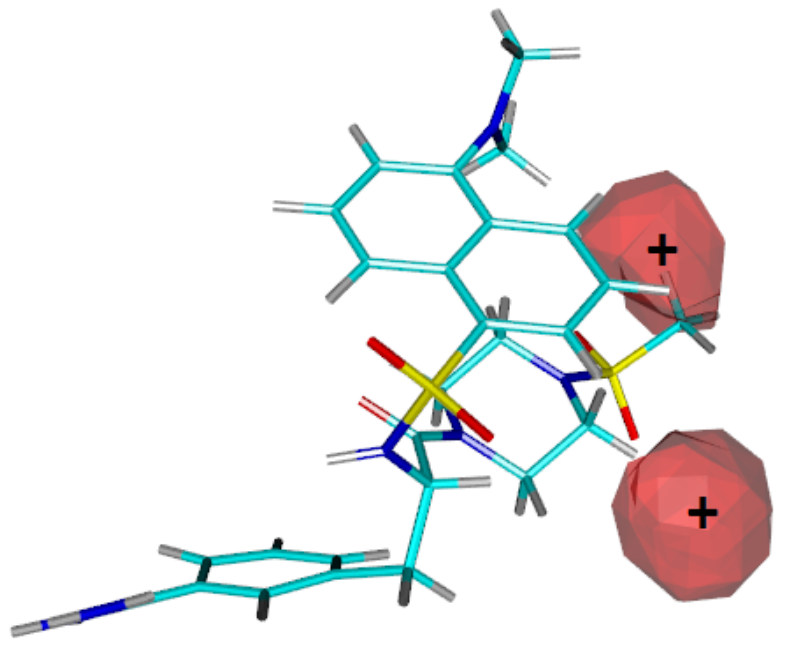 | Red color (marked with sign "+") means increase, blue color (marked with sign "-") means decrease of biological activity upon increase of partial electric positive charge on the nearest atoms |
| Steric | 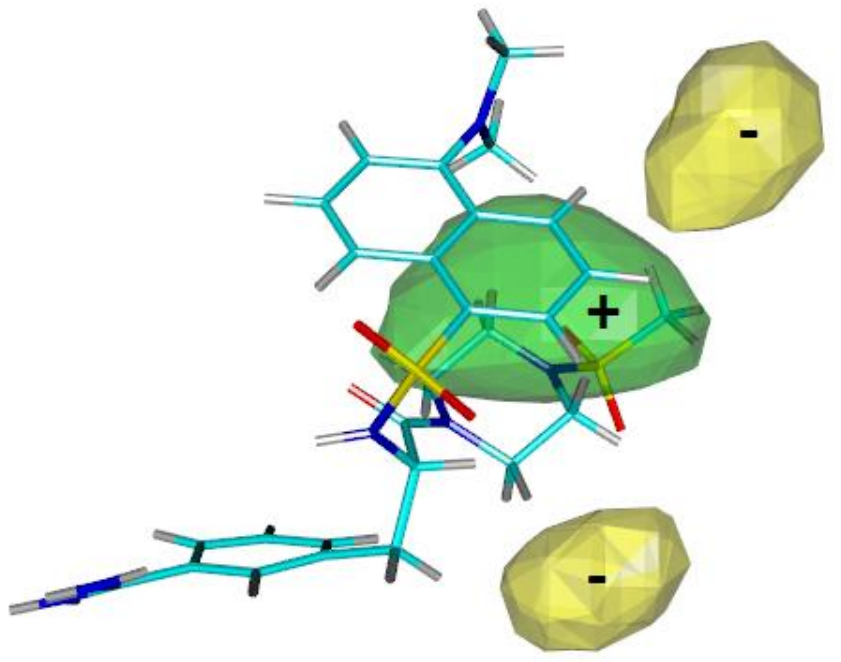 | Green color (marked with sign "+") means increase, yellow color (marked with sign "-") means decrease of biological activity upon increase of bulkiness of the nearest atoms. |
| Hydrophobic | 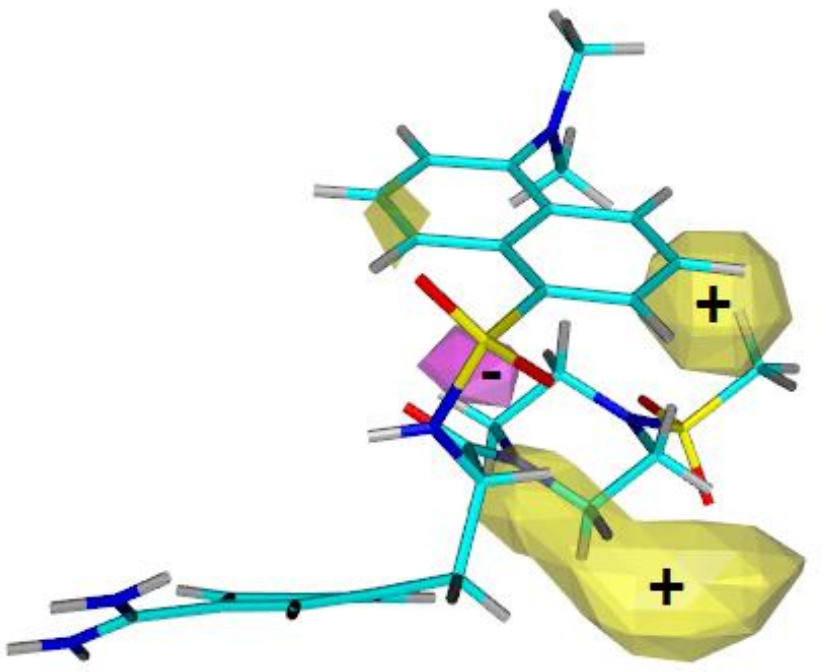 | Yellow color (marked with sign "+") means increase, violet color (marked with sign "-") means decrease of biological activity upon increase of lipophilicity of the nearest atoms. |

The benefit of such visualization for drug design is evident. Note, however, an important difference between isosurfaces in CMF models, from one side, and CoMFA, CoMSIA and GRID contour maps, from the other: the former are centered on atoms (as follows from Eq. 0.10), whereas the latter are situated around molecules (as a consequence of the impossibility to place probe atoms and groups inside atoms). Although such location of isosurfaces might seem unusual from the first glace, but it offers more direct answer to the question as to what changes should be introduced in order to increase biological activity of chemical compound.

Replacement of binding measures (such as $\log(1/IC_{50})$) in relation to individual targets for the difference in binding to two targets as an output of regression equation leads to the concept of selectivity fields, formulated by us earlier for CoMFA analysis [28]. In the framework of the continuous molecular field approach analogous fields of regression coefficients can also be built. Graphical visualization of their isosurfaces clearly indicates the changes that should be introduced into chemical structure in order to tune their selectivity towards different biological targets.

Note also that, due to the lack of discrete grids, molecular fields are treated in CMF as continuous functions with respect to spatial coordinates and therefore can be differentiated and integrated (even analytically, due to the use of Gaussian basis function). Easy differentiability of molecular field functions allows for applying very powerful although still unexplored in chemistry apparatus of data analysis and visualization offered by a newly emerged branch of mathematical statistics, functional data analysis [29], to visualize both molecular fields and fields of model coefficients. One should also add that, by applying the above-discussed methodology, in the framework of CMF, due to modularity of kernel-based approaches, it is possible to deduce fields of model coefficients not only for 3D-QSAR regression, but also for: (a) classification (in the case of Support Vector Machines they describe angular orientation of the hyperplane separating active from inactive compounds in the infinite-dimensional feature space); (b) novelty detection (or one-class classification [30], which can be used for virtual screening); (c) clustering.; (d) dimensionality reduction; etc. We hope that numerous methods of analyzing and visualizing chemical databases and SAR/QSAR/QSPR models offered by the use of continuous molecular fields would deepen insights into the nature of structure-activity relationships and facilitate drug design. This is one of the main directions of our current studies in this direction and a topic of future publications.

## 0.3 Modelling biological activity

We validated the CMF approach in two case studies and obtained preliminary results, which have been published as short communications [7,12]. The first one dealt with the use of the CMF to build 3D-QSAR regression models [7]. In the second case study [12], the performance of a new method for virtual screening of organic compounds based on the combination of the CMF methodology with the one-class SVM method (1-SVM) has been assessed. In both cases the CMF has not only proven its efficiency, but has also demonstrated some advantages compared to state-of-the-art approaches in chemoinformatics.

### *0.3.1. Building 3D-QSAR regression models*

We have tested the performance of the CMF approach in building 3D-QSAR regression models by using eight data sets. These sets were selected as expanded benchmark for 3D-QSAR methods as well as examples of various types of biologycal activity of organic ligands. We have studied 114 angiotensin converting enzyme (ACE) inhibitors [31], 111 acetylcholinesterase (AChE) inhibitors [32], 163 ligands for benzodiazepine receptors (BZR) [32], 322 cyclooxygenase-2 (COX-2) inhibitors [32], 397 dihydrofolatereductase (DHFR) inhibitors [32], 66 glycogen phosphorylase b (GPB) inhibitors [33], 76 thermolysin (THER) inhibitors [33], and 88 thrombine (THR) inhibitors [34].

All data on these sets were taken from the supplementary materials to Sutherland's paper [32]. They included chemical structures, activity values, splitting into the training and test sets, ionization states and conformations for all molecules, their spatial alignment and partial charges on atoms. As indicated in [25] ionization states of molecules had been prepared by deprotonating carbocyclic acids and phosphates and protonating non-aryl basic amines (except $NH_2$ groups that coordinate Zn in the ACE set), energy-minimizing the aligned molecules with MMFF94S force field in Sybyl was used for determination of atomic coordinates, scaled MNDO ESP-fit partial charges [35] had been calculated for all atoms with MOPAC 6.0, except that for the THER set all partial charges on atoms had been computed using the Gasteiger-Marsili method [36] as also implemented in Sybyl. Several characteristics of benchmarking data sets are presented in Table 0.2.

**Table 0.2.** QSAR DataSets

| Ligand data set | Training set | Test set | Activity ranging |
|---|---|---|---|
| Angiotensin converting enzyme (ACE) inhibitors | 76 | 38 | $pI_{50}$ 2.1 -9.9 |
| Acetylcholinesterase (AchE) inhibitors | 74 | 37 | $pI_{50}$ 4.3-9.5 |
| Ligands for benzodiazepine receptors (BZR)* | 98 | 49 | $pI_{50}$ 5.5-8.9 |
| Cyclooxygenase-2 (COX-2) inhibitors* | 188 | 94 | $pI_{50}$ 4.0-9.0 |
| Dihydrofolatereductase (DHFR) inhibitors* | 237 | 124 | $pI_{50}$ 3.3-9.8 |
| Glycogen phosphorylase b (GPB) inhibitors | 44 | 22 | pKi 1.3-6.8 |
| Thermolysin (THER) inhibitors | 51 | 25 | pKi 0.5-10.2 |
| Thrombine (THR) inhibitors | 59 | 29 | pKi 4.4-8.5 |

*In the BZR, COX-2, and DHFR data sets several compounds (16, 40, 36, respectively) were considered as inactive and not included in the training and the test sets [32].

Statistical characteristics of CMF models obtained for these data sets were compared with the same characteristics built for corresponning data sets using the common 3D-QSAR methods, CoMFA (Comparative Molecular Fields Analysis) [17] and CoMSIA (Comparative Molecular Similarity Index Analysis) [23], based on the use of molecular fields. Data on CoMFA and CoMSIA models were taken from Ref. [25].

Statistical parameters of CMF, CoMFA and CoMSIA models are shown in Table 0.3. They include the values of 4 statistical parameters: $q^2$ and $RMSE_{cv}$ characterizing internal predictive performance, $R^2_p$ and $RMSE_p$ – external predictive performance estimated using a single external validation set.

CoMSIA (in Ref. [25] – CoMSIA2) models are based on electrostatic and steric fields molecular fields and also involve contributions from the hydrophobic and two hydrogen-bonding molecular fields. All CMF models are based on the use of all afore-mentioned five types of molecular fields. All CoMFA and CoMSIA models were obtained by using a lattice with 2Å spacing expanding at least 4Å in each direction beyond aligned molecules. Only the most predictive CoMFA and CoMSIA models are included in the Table 0.3. In the course of building CMF models only two hyper-parameters, attenuation factor $\alpha_f$ (which was kept the same for all types of molecular fields in this study) and regularization coefficient $\gamma$, were optimized. In this study we used the same fixed value for all mixing coefficients, $h_f = 1$. The use of only two adjustable hyper-parameters provided satisfactory external predictive performance for all data sets. Although optimization of mixing coefficients $h_f$ always leads to sharp increase of $q^2$, in several cases parameter $q^2_{ex}$ for the external predictive performance becomes lower. This might happen because of the "model selection bias"

For comparing predictive ability of CMF models with those published in literature, all the data sets were split into the training and the test sets as specified in Table 0.2. The training sets were used for building 3D-QSAR models and for assessing their internal predictive performance using the 10-fold cross-validation procedure. The test sets were used for assessing the external predictive performance of the models.

**Table 0.3.** Statistical parameters of 3D-QSAR CMF, CoMFA and CoMSIA (in Ref. [32] – CoMSIA2) models

| | CMF | | | CoMFA* | | | CoMSIA** | | |
|---|---|---|---|---|---|---|---|---|---|
| | $q^2$ | $R^2_p$ | $RMSE_p$ | $q^2$ | $R^2_p$ | $RMSE_p$ | $q^2$ | $R^2_p$ | $RMSE_p$ |
| ACE | 0.72 | 0.65 | 1.24 | 0.68 | 0.49 | 1.54 | 0.66 | 0.49 | 1.53 |
| AChE | 0.58 | 0.64 | 0.77 | 0.52 | 0.47 | 0.95 | 0.49 | 0.44 | 0.98 |
| BZR | 0.40 | 0.51 | 0.79 | 0.65 | 0.00 | 0.97 | 0.45 | 0.12 | 0.91 |
| COX-2 | 0.57 | 0.14 | 1.23 | 0.49 | 0.29 | 1.24 | 0.57 | 0.37 | 1.17 |
| DHFR | 0.67 | 0.65 | 0.80 | 0.49 | 0.59 | 0.89 | 0.57 | 0.53 | 0.95 |
| GPB | 0.69 | 0.51 | 0.84 | 0.42 | 0.42 | 0.94 | 0.61 | 0.59 | 0.79 |
| THER | 0.60 | 0.31 | 1.86 | 0.52 | 0.54 | 1.59 | 0.51 | 0.53 | 1.60 |
| THR | 0.73 | 0.63 | 0.66 | 0.59 | 0.63 | 0.70 | 0.72 | 0.63 | 0.69 |

*PLS components: 3 (Ace, BZR), 4 (GPB, THER, THR), 5 (AChE, COX-2, DHFR)
** PLS components */Additional fields:* 2/hydro (Ace), 3/hydro (BZR, THER), 4/hydro (GPB ), 4/hydro+ H-bonding (AChE, COX-2, DHFR, THR)

As it is clear from Table 0.3, models built for 7 data sets by using the CMF approach almost in all cases show better internal (cross-validation) predictive performance (i.e., higher $q^2$) then the corresponding models obtained by the CoMFA and CoMSIA methods. The $q^2$ values of the CMF models obtained for the COX-2 and BZR data sets are, respectively, equal and lower as compared to the corresponding CoMSIA models.

There is also a moderate advantage in external predictive performance (estimated on external test sets using the parameters $R^2_p$ and $RMSE_p$) of the CMF models over the CoMFA, models for 5 data sets (ACE, AChE, BZR, DHFR, and GPB), and CoMSIA models for 4 data sets (ACE, AChE, BZR and DHFR).

One can notice that the performance of CMF is closer to that of the CoMSIA approach in comparison with CoMFA. This could be attributed to the fact that the mathematical form of Eq. 0.5 resembles expressions for similarity indices in CoMSIA. So, in spite of absolutely different underlying ideas, CoMSIA can formally be regarded as a discretized approximation of the current version of CMF, or, *vice versa*, CMF – as a continuous functional extension of CoMSIA. Therefore, the difference between the models produced by these methods might result from the effect of field discretization, different statistical procedure and parameterization of molecular fields.

Thus, the 3D-QSAR models obtained by CMF are comparable by the predictive ability with models built by means of such popular state-of-the-art approaches as CoMFA and CoMSIA. Moreover, in some cases, e.g. for data sets ACE, AChE, BZR and DHFR, the CMF approach is clearly advantageous.

Parameters of external predictive performance of 3D-QSAR CMF models ($q^2_{ex}$ and $RMSE_{cvex}$) estimated using external 5-fold cross-validation procedure are shown in the Table 0.4.

**Table 0.4.** Cross-validated external predictive performance of 3D-QSAR CMF models

| Parameters / Datasets | ACE | AChE | BZR | COX-2 | DHFR | GPB | THER | THR |
|---|---|---|---|---|---|---|---|---|
| $q^2_{ex}$ | 0.67 | 0.54 | 0.27 | 0.41 | 0.67 | 0.59 | 0.40 | 0.71 |
| $RMSE_{cvex}$ | 1.31 | 0.84 | 0.65 | 0.89 | 0.75 | 0.71 | 1.56 | 0.54 |

Following to Table 0.4, in all cases the value $q^2_{ex}$, which characterizes the external predictive performance, is lower than the value $q^2$ computed using the internal cross-validation. This means that the use of only two adjustable hyperparameters may cause the “model selection bias”. Almost in all cases the value $q^2_{ex}$ lies between $R^2_p$ and $q^2$. It is interesting to note that $q^2_{ex}$ for CMF models are usually higher than $R^2_p$ for CoMFA and CoMSIA models. The predictive performance assessed using the external 5-fold cross-validation procedure is especially high for ACE, DHFR and THR.

### *0.3.2 Virtual Screening via Combination CMF with 1-SVM Technique*

As it follows from our earlier publications, one-class classification (novelty detection) machine learning methods is a mathematical base of a new general approach to conducting ligand-based virtual screening of chemical compounds [37,38]. Although several dozens of different algorithms for building one-class classification (novelty detection) models are known [39,40], only those of them which are based on using kernels are suited for working with continuous molecular fields. The One-Class Support Vector Machines (1-SVM) [11] machine learning method is one of them. It builds one-class classification models by seeking for hyperplane in infinite-dimensional functional Hilbert space (feature space) with maximum distance from the coordinate origin and separating a given proportion of training examples from it. In this case, the field of model coefficients is formed by leading cosines of the normal to this hyperplane. In the physical space they form description of “ideal” molecular fields (shapes), which are compared with molecular fields of test molecules. This “ideal” combination of fields reflects the structure of the corresponding binding site in biological macromolecule and can be viewed as a “negative image” of its molecular fields (see Fig. 0.1).

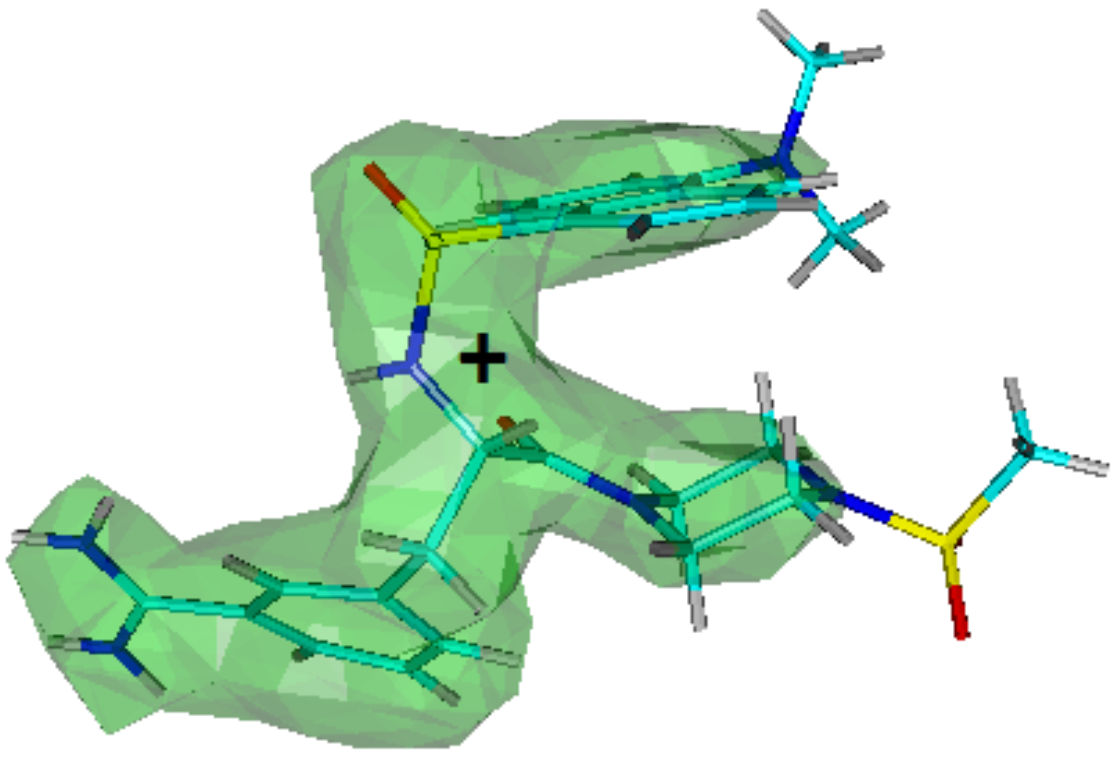

**Fig. 0.1**. The “ideal” steric molecular field corresponding vector perpendicular to separating hyperplane in the 1-SVM model built for thrombin inhibitors (2-amidinophenylalanines). Its isosurface can be viewed as a “negative image” of the binding site in biological target.

So, 1-SVM models in conjunction with molecular field kernels perform ligand-based virtual screening based on similarity of molecular fields (shapes). In comparison with other shape-based similarity search methods, they are much more flexible, because one can find the optimal degree of generalization (simplification). See Fig. 0.2, in which the level of generalization of model coefficient field description increases from the lower to the upper row. This leads to high performance in ligand-based virtual screening [12].

| Field | Electrostatic | Steric | Hydrophobic |
|---|---|---|---|
| $\alpha_f = 0.02$ | 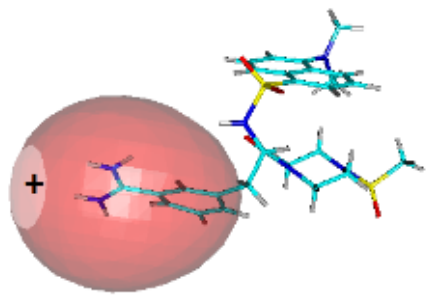 | 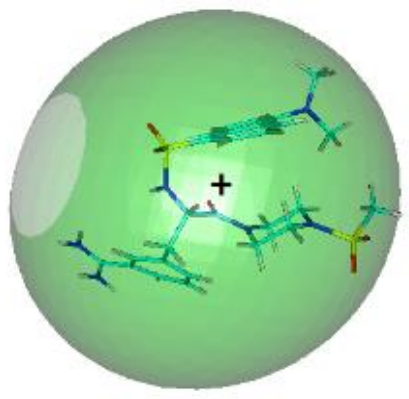 | 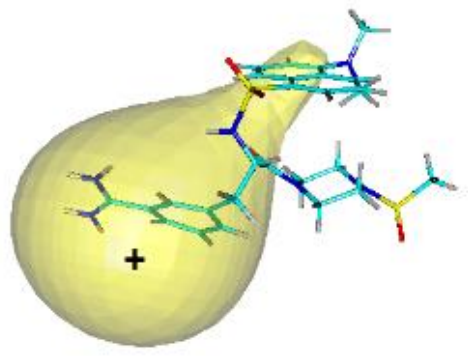 |
| $\alpha_f = 1.0$ | 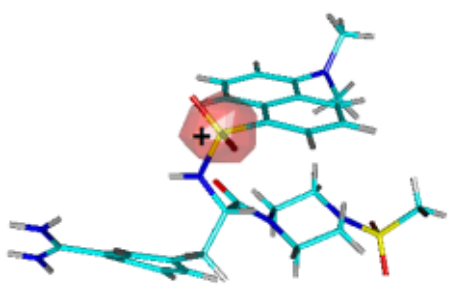 | 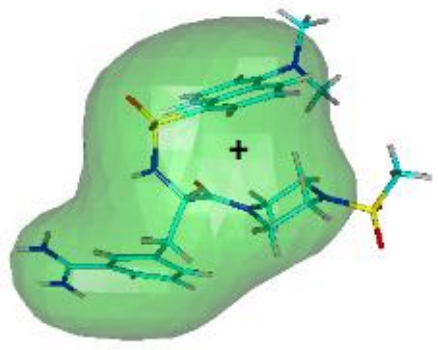 | 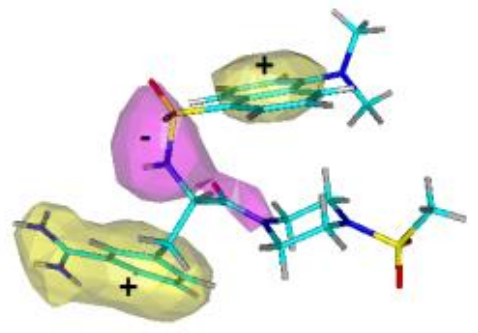 |

$\alpha_f = 5.0$

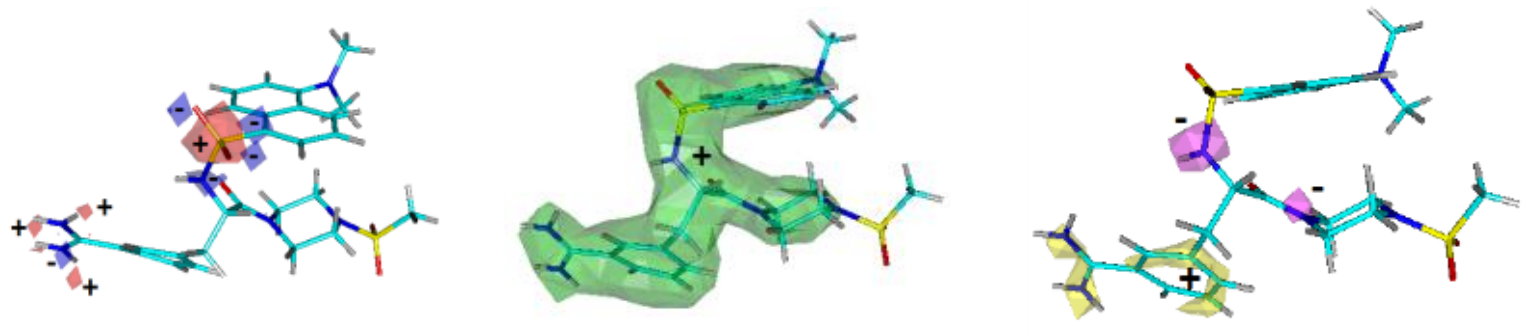

**Fig. 0.2.** Isosurfaces for the fields of 1-SVM model coefficients for thrombin inhibitors (2-amidinophenylalanines).

In the following case study [12], we have assessed the performance of a new promising method for virtual screening of organic compounds based on a combination of the CMF methodology with the one-class SVM method (1-SVM). The first step of constructing the model was spatial alignment of the structures of organic ligands. In this work, the alignment was performed with the SEAL algorithm [41] implemented by us in the framework of the software for CMF modeling. Then, kernel values were calculated, and the model was constructed using the LibSVM program [42].

In the one-class classification method, only active structures are used. Sequentially excluding one structure at a time and constructing the model based on the remaining structures, one can predict the activity of all active compounds. However, for assessing the statistical characteristics of classification models, it is necessary to determine not only the number of active compounds predicted to be active (true positive, TP) and the number of active compounds predicted to be inactive (false negative, FN) but also the number of inactive compounds predicted to be inactive (true negative, TN) and the number of inactive compounds predicted to be active (false positive, FP). To determine the last two characteristics, we used structures resembling the structures of active ligands in their physicochemical properties but presumed to be inactive (so-called decoys).

We have built one-class models for data taken from the DUD database [43], which contains the structures of active ligands for different biological targets, as well as the structures of corresponding decoys. It is worth noting that the latter were used only for assessing the statistical characteristics of classification models and were not involved in their construction. In particular, decoys were used for determining the TN and FP models constructed with the use of active compounds.

The suggested one-class classifier calculates a continuous quantity (a classifier function), for which the threshold value is determined. If the classifier function calculated for a certain ligand exceeds the threshold value, the compound is considered active; otherwise, the structure is discarded from further consideration.

Dependence of the FN, FP, TN, and TP on the threshold value is clearly reflected by a Receiver Operator Characteristic (ROC) curve [44] in the TPR–FPR coordinates (true positive rate versus false positive rate), where TRP = TP/(TP +

TN) and FPR = FP/(FP + TN). The larger the area under the curve (AUC), the higher is the classifier efficiency.

Constructing the classification model necessitates maximizing the AUC value by optimizing the 1-SVM parameter ν and parameter $\alpha_f$ of the CMF kernel from Eq. 0.5. We have studied both the individual electrostatic, steric, and hydrophobic kernels and their linear combinations. For individual kernels, two parameters have been optimized: $v$ and the $\alpha_f$ parameter corresponding to a given type of molecular field. At the first step, the optimization algorithm was launched ten times, each time starting from a set of random parameter values in the ranges ν ∈ [0.01; 0.80], $h_f$ ∈ [0.0001;0.3000], and $\alpha_f$ ∈ [0.001; 1.000]. At the second step, the Nelder–Mead algorithm was used for refining the optimal parameters; the set of the best-fit parameters obtained at the first step of optimization were used as the initial approximation.

Tables 0.5 and 0.6 summarize the results of building 1-SVM models on the basis of continuous molecular fields for HIV reverse transcriptase (HIVRT) and trypsin inhibitors. As follows from Table 0.5, the best performance for HIVRT is obtained by the model constructed using the steric kernel and resulting in an AUC value of 0.75. For this target, the use of a linear combination of several kernels does not improve the AUC value. At the same time, for trypsin inhibitors, rather high AUC values (0.86–0.91) were obtained on the basis of individual models constructed with the use of all three kernels, which is likely due to their mutual correlation. However, for this target, the use of a linear combination of all kernels increases the AUC value up to 0.94.

**Table 0.5.** Parameters and AUC of the models for HIV reverse transcriptase inhibitors obtained by the 1-SVM method with the use of kernels in the framework of the CMF

| Molecular field | $v$ | $\alpha_{el}$ | $\alpha_{st}$ | $\alpha_{hyd}$ | AUC |
|---|---|---|---|---|---|
| Electrostatic | 0.082 | 0.031 | - | - | 0.60 |
| Steric | 0.002 | - | 0.010 | - | 0.75 |
| Hydrophobic | 0.466 | - | - | 0.009 | 0.65 |

**Table 0.6.** Parameters and AUC of the models for trypsin inhibitors obtained by the 1-SVM method with the use of kernels in the framework of the CMF

| Molecular Field | $v$ | $h_{el}$ | $\alpha_{el}$ | $h_{st}$ | $\alpha_{st}$ | $h_{hyd}$ | $\alpha_{hyd}$ | AUC |
|---|---|---|---|---|---|---|---|---|
| Electrostatic | 0.53 | - | 0.30 | - | - | - | - | 0.91 |
| Steric | 0.47 | - | - | - | 0.00 | - | - | 0.87 |
| Hydrophobic | 0.66 | - | - | - | - | - | 0.30 | 0.86 |
| Linear combination of | 0.45 | 0.32 | 0.30 | 0.58 | 0.04 | 0.11 | 0.15 | 0.94 |

| the fields |
|---|

Our results show of the effectiveness of the suggested methodology. Being an alternative to the traditional similarity search for active compounds using of the Tanimoto coefficient and to the binary classification methods, this approach possesses unique properties. In contrast to the binary classification methods, the described method is not sensitive to the choice of counterexamples. As distinct from the traditional similarity search, the suggested method can adapt to complex structure–activity landscapes and, thus, makes it possible to avoid activity cliffs [45]. In addition, as compared to similarity search methods based on the use of fragmental descriptors, the suggested approach implies using the same model to the sets of compounds belonging to different structural classes.

## 0.4 The Main Directions of Further Development of the CMF Approach

In this article, we have considered a particular implementation of the CMF approach aimed at building 3D-QSAR models. This implementation is being actively developed, and its future version will surely be better than the current one. This section, however, concerns more global, strategic directions of further development of the whole CMF approach.

### *0.4.1 Introduction of Additional types of Molecular Fields. Integration with Quantum Chemistry*

The CMF approach is not confined to the simplest approximation scheme introduced by Eq. 0.5. Any number of Gaussian functions, (both isotropic, i.e. spherically symmetrical, and non-isotropic) as well as any other set of basic functions (such as splines, wavelets, etc) can be used for approximating continuous molecular fields. This provides the ability to work with complex types of molecular fields, including those derived from the electron density function.

Close integration with quantum chemistry is a promising direction of further development of the CMF approach. Indeed, wave and electron density functions are the most natural, comprehensive and accurate way to describe molecules. Moreover, all possible continuous molecular fields can be generated from the electron density functions with the help of integral transforms [46]. Especially promising is the use of conceptual DFT molecular fields [47] based on conceptual DFT [48]. Molecular fields (steric, electrostatic, local softness and LUMO) derived from the electron density function have already been successfully used for build-ing 3D-QSAR models in the field of metal complex catalysis using the traditional grid-based CoMFA approach [49,50]. We can expect that due to the ability to approximate all kinds of molecular fields with any degree of accuracy the CMF approach will be very useful in the development of new catalysts and supramolecular complexes, which require high accuracy in representing the electronic structure of molecules.

### *0.4.2 The Issue of Molecular Alignment*

All methods of molecular alignment useful for building traditional lattice-based 3D-QSAR models can also be applied in the framework of the CMF approach. Meanwhile, thanks to the integrability of continuous functions describing molecular fields, the latter approach offers additional possibilities.

Molecular alignment exactly corresponds to the "curve registration" procedure, which is the first necessary step in any functional data analysis [29]. In the case of biological activity caused by protein-ligand interactions, mutual orientation of different ligands inside binding sites defines their "natural alignment". Such alignment can be checked by means of independent physical experiments (X-ray diffraction, NMR, etc), and hence its explicit consideration is equivalent to the inclusion of "external domain knowledge", which is always preferred in machine learning [4]. Therefore, the choice of the strategy for molecular alignment should be governed by the necessity to mimic mutual orientation of molecules in the underlying physical processes. It should however be pointed out that that this does not mean that the "natural alignment" should always provide the strongest 3D-QSAR models.

In the case when the exact information concerning the structure of binding pockets of biological macromolecules is not available, the CMF approach provides a consistent criterion for the pairwise alignment of molecules $i$ and $j$: maximization of the kernel $K(M_i, M_j)$. Indeed, its form appears to be closely related to the function used in the SEAL program for aligning molecules [41]. So, the CMF approach offers the possibility to use the same function both for aligning molecules and building structure-activity models. This could lead to more close integration of molecular alignment into the process of chemical data analysis. Moreover, the CMF approach provides an additional criterion for the multiple alignment of molecules – the "compressibility" of molecular fields, which can be assessed using unsupervised dimensionality reduction approaches, such as the kernel (functional) principal component analysis. Both criteria can also be applied to choose molecular conformations for tackling the problem of molecular flexibility.

The issue of alignment-free approaches deserves special attention. Sometimes the necessity to perform alignment of molecules in several grid-based 3D-QSAR methods, such as CoMFA and CoMSIA, is considered as limitation of such approaches, which should be avoided [51]. This has led to the development of alignment-free approaches, such as those based on autocorrelation vectors [52], molecular moments (CoMMA) [53], 3D WHIM [54], EVA [51], GRIND [55], FLAP [56], VolSurf [57] descriptors, etc. The CMF approach provides a new kind of solutions based on the use of 3D-rotation invariant kernels [58,59] in the frame of the concept of invariant pattern recognition [60]. The feasibility of this approach results from the ability to apply the required integral transforms to continuous functions describing molecular fields.

### *0.4.3 Taking into account molecular flexibility*

A universal way of tackling the problem of molecular flexibility was suggested in paper [61] for kernel-based methods. It consists in averaging kernels over all conformations for each molecule. This approach is however computationally feasible only for very small number of conformations per molecule. Indeed, consideration of only 20 conformations for each molecule results in the necessity to consider 400 pairs of conformations in order to fill each cell in kernel matrix. In addition to huge computational burden, this approach does not solve the problem of alignment and therefore applicable only for alignment-free approaches.

The CMF approach can offer an alternative solution to this problem. Instead of using discrete sets of "representative" conformations, one can consider for each molecule an infinite number of conformations organized into a continuous manifold, so-called "conformational space". This provides the ability to apply functional data analysis not only to molecular fields but also to molecular geometry in a consistent way. Such "conformational space" can be described by means of some probability density function (*pdf*) in 3N-dimensional Euclidean space, where N is the number of atoms in the molecule under study. Having applied several approximations from the arsenal of statistical physics, one can obtain the following expression for calculating atomic kernels instead of Eq. 0.6:

$$k_f(A_{il}, A_{jm}) = w_{fil} w_{fjm} \sqrt{\frac{\pi^3}{(\alpha_f)^3}} \int_{\Re^3}\int_{\Re^3} \exp(-\frac{\alpha_f}{4}\left\|\mathbf{r}_{il} - \mathbf{r}_{jm}\right\|^2) p(\mathbf{r}_{il}) p(\mathbf{r}_{jm}) d\mathbf{r}_{il} d\mathbf{r}_{jm} \qquad (0.11)$$

where $p(\mathbf{r}_{il})$ is the one-particle *pdf* for atom *l* in molecule *i*. In order to compute all necessary *pdf*, one should perform molecular dynamics or Monte-Carlo studies for all molecules, collect conformations along trajectories, align them (e.g., using the common template), and apply the GMM algorithm [62] to approximate *pdf* for each atom as a mixture of several Gaussian functions. In this case the integral in Eq. 0.11 contains the product of Gaussian functions and hence can be computed analytically. Therefore, due to easy integrability of continuous molecular field functions, it is possible to build models taking into account the whole „conformational spaces“. For modeling receptor-ligand interactions, it is important to per-form molecular dynamics simulations of ligands either inside the binding pockets or using their simplified surrogates. In the simplest case this amounts to choosing a single „biologically active“ conformation from the results of molecular docking.

### *0.4.4. Prediction of physico-chemical properties*

It is expected that the CMF approach will be extended to predict physico-chemical properties of chemical compounds and their supramolecular complexes. The theoretical possibility of this follows from the following analysis.

Due to the ability to apply methods of functional analysis, continuous molecular fields can be tailored for solving many different tasks in chemoinformatics. Consider, for example, prediction of physic-chemical properties in diverse datasets. In this case, "natural alignment" corresponds to the uniform probability of molecules to adopt any possible mutual orientation. Therefore, kernel $K_f(\mathrm{M}_i, \mathrm{M}_j)$ describing the similarity between the molecular fields of the $f$th type for the $i$th and $j$th molecules can be computed by averaging over all possible mutual orientations:

$$K_f(M_i, M_j) = \sum_l \sum_m \iiint k_f(A_{il}, A_{jm}) d^3\mathbf{r}_{jm} = \frac{8\pi^3}{(\alpha_f)^3} \sum_l \sum_m w_{fil} w_{fjm} \quad (0.12)$$

The resulting kernel does not depend on molecular geometry, although it includes all constants describing continuous molecular fields. This kernel can be considered as a particular case of the molecular convolution kernels for which the ability to model additive physicochemical properties was shown by us earlier [63]. Extension to the case of modeling metal complexation would require integration over only two angles $\theta$ and $\varphi$ of spherical coordinate system, whereas for modeling cyclodextrine complexation it is sufficient to integrate over the single angle $\varphi$ of the cylindric coordinate system, as follows from consideration of the "natural alignment" in each of these systems. One can show that in these cases all necessary integrals can be taken in analytical form using Bessel functions. Such universality and flexibility results from integrability of continuous molecular fields. This opens a direct way to extending the CMF approach to predicting physico-chemical properties of chemical compounds.

### *0.4.5 Taking into account different ionization states, tautomers and conformers*

The CMF approach can be extended to the case of the existence of several tautomers, ionization (protonation) states and conformers by replacing the Eq. 0.5 with its more general form:

$$\rho_{fil}(\mathbf{r}) = \sum_s v_s \sum_t v_{st} \sum_c v_{stc} w_{fil} \exp(-\frac{1}{2}\alpha_f \|\mathbf{r} - \mathbf{r}_{il}\|^2), \qquad (0.13)$$

where index $s$ counts different ionization (protonation) states, index $t$ counts tautomers, index $c$ counts conformers, $v_s$ is the population of the ionization state $s$, $v_{st}$ is the relative population of tautomer $t$ in ionization state $s$, while $v_{stc}$ is the relative population of the conformer c of the molecule in ionization state $s$ and in tautomeric state $t$. Populations $v_s$, $v_{st}$ and $v_{stc}$ can be assessed using molecular modeling simulations.

### *0.4.6 Further extension of the approach to encompass biological macromolecules and their interactions with both macro- and small molecules*

One of the most promising ways to develop further the CMF approach is its further extension to the description of the binding sites of biomolecules and their interactions with ligands. This seems to be feasible, because molecular fields of biological targets are identical by their nature to those of small ligands. Because of this, many of the approaches and methods originally developed for working with small molecules can be transferred to biological macromolecules.

One can suggest several ways of conducting research in this direction. First, kernels for comparing molecular fields of biological targets (or their binding sites) could be constructed in exactly the same way as it has been done small molecules and described in this paper. By combining them with various kernel machine learning methods, various regression, classification, novelty detection, ranking and dimensionality reductions tasks could be formulated and solved for biological macromolecules and their binding sites. This could lead to promising applications in biology-related sciences.

Second, kernels for protein-ligand pairs can be constructed by combining kernels for small molecules (ligands) and kernels for macromolecules (proteins), as it was done by Erhan *et al* [64], Faulon *et al* [65], Jacob and Vert [66], and Bajorath *et al* [67]. The main advantage of using continuous molecular fields in this case is that this approach can be applied consistently to construct kernels for both small organic and big biological macromolecules using the same typed of molecular fields. Thanks to this, the same data analysis methods could be used to describe also protein-peptide, protein-protein, protein-DNA, protein-RNA interactions, as well as properties of peptides and proteins with non-standard residues.

Third, kernels for protein-ligand interactions can be constructed in the frame of the CMF approach by encapsulating products of molecular fields of ligand and protein into kernels. Such combined kernels could easily be used in conjunction with various kernel machine learning methods to solve different task relating to protein-ligand interactions.

It is expected that the results obtained in this direction might be useful in the field of chemogenomics for target profiling, in bioinformatics and proteomics for classifying and annotating protein macromolecules and their binding sites by considering and comparing their molecular fields, and in system biology for predicting interaction graphs for biomolecules, and in drug design.

### *0.4.7 Integration of approaches to prevent the model selection bias*

As it has already been discussed, one of the drawbacks of the CMF approach in its present state is the danger of over-fitting because of the model selection bias [68]. Several ways to prevent this phenomenon have been suggested in literature, including Bayesian regularization of hyper-parameters [69] and hyper-parameter averaging [70]. Algorithms of multiple kernel learning [71] might also be useful in this case. For small datasets the preferred solution would be to apply the full Bayesian approach [62], where the hyper-parameters are integrated out rather than optimized. Integration of approaches to prevent the model selection bias is one of the most important tasks for further development of the CMF approach.

### *0.4.8 Combining with different machine learning methods*

The CMF approach is easily extensible tanks to its modularity. By combining different types of molecular fields, different types of kernels with different types of kernel-based machine learning methods, one can obtain various methods for building SAR/QSAR/QSPR models and conducting virtual screening. Although some of such methods may be similar to the existing ones, nonetheless it is likely that some of them will be fundamentally novel approaches. Table 0.7 lists different tasks being solved by kernel-based machine learning methods, the names of such methods, and the roles that the CMF approach could play in conjunction with them in chemoinformatics. The first row in this table deals with the regression task considered in this paper. The second row concerns the use of molecular kernels in combination with one-class classification kernel-based methods for conducting virtual screening based on the similarity of molecular fields. The feasibility of this approach has already been proved by us, see preliminary communication [12]. The rest of the table shows the promising directions for further development of the CMF approach.

**Table 0.7.** The use of CMF in conjunction with different kernel-based machine learning methods

| Machine Learning Task | Machine Learning Methods | Role in Chemoinformatics |
|---|---|---|
| Regression | Support Vector Regression (SVR) [5], Kernel Ridge Regression (KRR) [9], Kernel Partial Least Squares (KPLS) [27], Gaussian Processes for Regression (GP-R) [10] | QSAR/QSPR |
| One-class classification (novelty detection) | One-Class Support Vector Machine (1-SVM) [11], Support Vector Data Description (SVDD) [72] | Virtual screening based of similarity of molecular fields |
| Binary and multi-class classification | Support Vector Machines (SVM) [8], Gaussian Processes for Classification (GP-C) [10] | Classification of chemical compounds (active/inactive), predicting profiles of biological activity for chemical compounds |
| Dimensionality reduction and data visualization | Kernel Principal Component Analysis (KPCA) [6], Kernel Feature Analysis (KFA) [73] | Drawing maps of chemical space |
| Cluster analysis | Kernel k-means [74] | Classification of chemical compounds by mechanism of action (including binding mode) |
| Canonical correlation | Kernel Canonical Correlation Analysis (KCCA) [75] | Relationships between molecular fields of ligands and molecular fields of their binding sites |

## 0.4 Conclusion

The CMF approach describes molecules by ensemble of continuous functions (molecular fields), instead of finite sets of molecular descriptors (such as interaction energies computed at grid nodes). The potential advantages of this approach results from the ability to approximate electronic molecular structures with any desirable accuracy level, the ability to leverage the valuable information contained in partial derivatives of molecular fields (otherwise lost upon discretization) to analyze models and enhance their predictive performance, the ability to apply integral transforms to molecular fields and models, etc.

The most attractive features of the CMF approach are its versatility and universality. By combining different types of molecular fields and methods of their approximation, different types of kernels with different types of kernel-based machine learning methods, it is possible to present lots of existing methods in chemoinformatics and medicinal chemistry as particular cases within a universal methodology. The CMF methodology can easily be extended to building classification and novelty detection models, visualizing them, performing virtual screening, processing diverse datasets.

We see the following main directions for further development of the CMF approach: introduction of additional types of molecular fields, including conceptual DFT molecular fields; tackling the issue of molecular alignment and flexibility; taking into account molecular flexibility; prediction of physico-chemical properties of chemical compounds; taking into account different ionization states of molecules, their tautomers and conformers; extension of the approach to work with biological macromolecules and supramolecular complexes; integration of special approaches for preventing model selection bias; combining with different machine learning methods aimed at solving various tasks.

**Acknowledgments.** The authors thank Prof. Yu.A.Ustynyuk for stimulating discussion and advice. The authors also thank Prof. A.Varnek and Dr. G.Marcou for valuable comments regarding the developed approach. This work was supported by Russian Foundation for Basic Research (Grant 13-07-00511).